\documentclass[aps,prb,twocolumn]{revtex4}
\usepackage{graphics}
\usepackage{graphicx}
\usepackage{array}
\usepackage{dcolumn}
\usepackage{bm}
\usepackage{float}
\usepackage{enumitem}
\usepackage[colorinlistoftodos, textwidth=4cm, shadow]{todonotes}

\begin{document}

\title{Effect of Eu magnetism on the electronic properties of the candidate Dirac material EuMnBi$_2$}

\author{Andrew F. May}
\email{mayaf@ornl.gov}
\author{Michael A. McGuire}
\author{Brian C. Sales}
\affiliation{Materials Science and Technology Division, Oak Ridge National Laboratory, Oak Ridge, TN 37831}

\date{\today}

\begin{abstract}
The crystal structure and physical properties of the layered material EuMnBi$_2$ have been characterized by measurements on single crystals.  EuMnBi$_2$ is isostructural with the Dirac material SrMnBi$_2$ based on single crystal x-ray diffraction, crystallizing in the $I4/mmm$ space group (No. 139).  Magnetic susceptibility measurements suggest antiferromagnetic (AFM) ordering of moments on divalent Eu ions near $T_N=22$\,K.  For low fields, the ordered Eu moments are aligned along the \textit{c}-axis, and a spin-flop is observed near 5.4\,T at 5\,K.  The moment is not saturated in an applied field of 13\,T at 5\,K, which is uncommon for compounds containing Eu$^{2+}$.  The magnetic behavior suggests an anisotropy enhancement via interaction between Eu and the Mn moments that appear to be order antiferromagnetically below $\approx$310\,K.  A large increase in the magnetoresistance is observed across the spin-flop, with absolute magnetoresistance reaching $\approx$650\% at 5\,K and 12\,T.  Hall effect measurements reveal a decrease in the carrier density below $T_N$, which implies a manipulation of the Fermi surface by magnetism on the sites surrounding the Bi square nets that lead to Dirac cones in this family of materials.
\end{abstract}

\maketitle

\section{Introduction}

Layered pnictides are prevalent in a variety of technologically-important research fields that are currently of great interest to many condensed matter researchers.  The observation of superconductivity in iron-containing pnictides, such as those derived from the nominally tetragonal BaFe$_2$As$_2$ and LaFeAsO, has led to a significant interest in the structure-property relationships of these large families of layered materials.\cite{SinghReview2009,Mandrus2010,JohnstonReview2010,PaglioneGreen2010}  Also, ternary pnictides with the trigonal CaAl$_2$Si$_2$ structure-type possess promising thermoelectric efficiency.\cite{122Jeff,Cd122Grin09,YbCdMnSb2,122Tober} Furthermore, three-dimensional architectures that are crystallographically layered are of interest as bulk-sources for two-dimensional (2D) electronic or magnetic materials.\cite{2D_review2013,2D_calc_PRX2013}

EuMnBi$_2$, which is isostructural with SrMnBi$_2$, is structurally-related to the ferrous pnictide superconductors.  In BaFe$_2$As$_2$, for instance, an FeAs layer composed of edge-sharing FeAs$_4$ tetrahedra is supported by nets of Ba ions.  EuMnBi$_2$ is similarly built from layers composed of edge-sharing MnBi$_4$ and Eu nets, as shown in the graphic within Table \ref{tab:refine}.  However, the stacking sequence is complicated by the presence of layers of Bi atoms in a square-net configuration, which results in a significantly larger unit cell along the $c$-axis.  Beyond a structural modification, these cation-capped bismuth layers have a significant influence on the electronic transport properties.\cite{Wang2011}

Dirac fermions have been reported in SrMnBi$_2$, and the structurally-similar compounds CaMnBi$_2$ and LaAgBi$_2$/LaAgSb$_2$, on the basis of first principles calculations, analysis of physical property data, and angle-resolved photoemission spectroscopy (ARPES).\cite{Wang2011,Park2011PRL,Wang2012PRB,He2012APL,Lee2013PRB,Wang2013PRB,Feng2013}  Linear bands associated with Dirac points near the Fermi energy make these compounds of interest for fundamental studies, as well as possible electronic applications. For instance, these linear bands have a large magnetoresistance that increases linearly with field and does not saturate.\cite{He2012APL,Wang2013PRB,Abrikosov2003} Unlike in graphene, the Dirac cones are anisotropic in SrMnBi$_2$ and perhaps also in CaMnBi$_2$, with ARPES measurements reporting both to be anisotropic while first principles calculations suggest only SrMnBi$_2$ should possess anisotropic Dirac cones due to the different stacking of Ca/Sr around the Bi nets.\cite{Lee2013PRB,Feng2013} 

In SrMnBi$_2$, the Mn moments order antiferromagnetically below approximately 280\,K, with moments aligned along the $c$-axis.\cite{Park2011PRL} and similar ordering is observed below approximately 270\,K in CaMnBi$_2$.\cite{He2012APL}  Recent work on LaAgBi$_2$, however, has shown that magnetism on the Mn site is not critical for the interesting electronic behavior.\cite{Wang2013PRB} First principles calculations suggest that the Bi nets and their interactions with the neighboring cation (Ca/Sr/La) dominate the electronic dispersion at the Fermi level (or in the very least the electronic properties).\cite{Wang2011,Lee2013PRB}  For this reason, we investigated the compound EuMnBi$_2$, which possesses divalent Eu ions that are characterized by a large J=$\frac{7}{2}$ but have L=0. This material may provide a platform to gain unique insights into the manipulation of Dirac fermions in this class of ternary pnictides, and thus the properties reported may also serve as a starting point for future theoretical investigations.

We report the crystal structure and physical properties of EuMnBi$_2$ from measurements on single crystals.  High quality structural refinements of single crystal x-ray diffraction data were obtained using space group \textit{I4/mmm} (139), and the SrZnBi$_2$ (SrMnBi$_2$) structure type.  The magnetization data and an anomaly in the specific heat capacity are consistent with moments on divalent Eu ions ordering antiferromagnetically near 22\,K, and a simultaneous decrease in the carrier density is inferred from Hall effect data.  In the low-temperature ordered state, a spin-flop is observed near 5.4\,T for fields applied along the $c$-axis.  In addition, the magnetic moment is not saturated at 5\,K in an applied field of 13\,T.  Both of these magnetic features are atypical of layered compounds containing divalent europium.  This suggests an interaction between Mn and Eu moments, the former of which likely order antiferromagnetically near room temperature.  Also, the large and positive magnetoresistance (MR) responds strongly to the spin-flop, with a substantial increase that mimics the increase in the measured moment.

\section{Experimental Details and Crystallographic Data} 

Single crystals were grown from a bismuth-rich melt in alumina crucibles sealed under vacuum inside fused quartz ampoules.  High purity Bi, Eu (Ames Lab), and Mn were combined in 2 or 5\,mL Al$_2$O$_3$ crucibles inside a helium glove box at the stoichiometry EuMnBi$_4$.  Crucibles filled with quartz wool were placed on top of the growth crucible to collect the discarded flux.  During heating, an initial dwell of 4\,h at 500\,$^{\circ}$C was followed by a 20-24h soak at a minimum of 750$^{\circ}$C. Relatively large crystals were obtained for growth rates between 1.5 and 4$^{\circ}$/h, with excess flux removed at 525-600$^{\circ}$C by centrifugation.  Single crystals of SrMnBi$_2$ were prepared in a similar manner, with the primary purpose of obtaining a phonon-background for the analysis of the low-temperature specific heat anomaly in EuMnBi$_2$.

The as-grown crystals were often larger than 5\,mm on a facet, which were confirmed to be normal to the $c$-axis by x-ray diffraction.  Due to the plate-like nature of these crystals, as well as their malleability, it was difficult to isolate a crystal appropriate for single crystal x-ray diffraction.  Therefore, crystals for single-crystal x-ray diffraction were selected from a sample prepared by melting a near-stoichiometric (slightly Eu rich) mixture of the elements at 1050$^{\circ}$C in a vacuum-sealed Al$_2$O$_3$ crucible, followed by cooling at 100$^{\circ}$/h.  The crystal structure was solved from single crystal x-ray diffraction obtained at 173\,K.  The data were collected using Mo-$K\alpha$ radiation ($\lambda$ = 0.71073\,\AA) on a Bruker SMART APEX CCD.  During refinement of the crystal structure, absorption corrections were applied with SADABS and the data were refined using SHELXL-97.\cite{Shelxl97}  2338 reflections were measured, giving 214 unique reflections (12 refinement parameters).  The Goodness of fit was 1.292 while R$_{\mathrm{int}}$ = 0.0304.

EuMnBi$_2$ is found to be isostructural to SrMnBi$_2$ (SrZnBi$_2$ structure type), with tetragonal unit cell dimensions of $a$=4.5342(6)\,\AA\, and $c$=22.427(4)\,\AA\, at approximately 173\,K.  The pertinent structural data are summarized in Table \ref{tab:refine}.  There are two free atomic positions in this structure, for Eu and Bi2, and the data reported in Table \ref{tab:refine} are similar to those reported for SrMnBi$_2$.\cite{Cordier1977}

\begin{table}[h!]
\caption{Crystal structure and selected results from refinement of single crystal diffraction data for EuMnBi$_2$ collected at $T$$\approx$173\,K (space group $I$4/mmm, No. 139).}
  \label{tab:refine}
  \centering
  \begin{tabular}{  m{1.85cm} m{3.5cm}  m{2.9cm} }
    \hline
     \begin{minipage}{0.32\columnwidth}
    \includegraphics[width=\columnwidth]{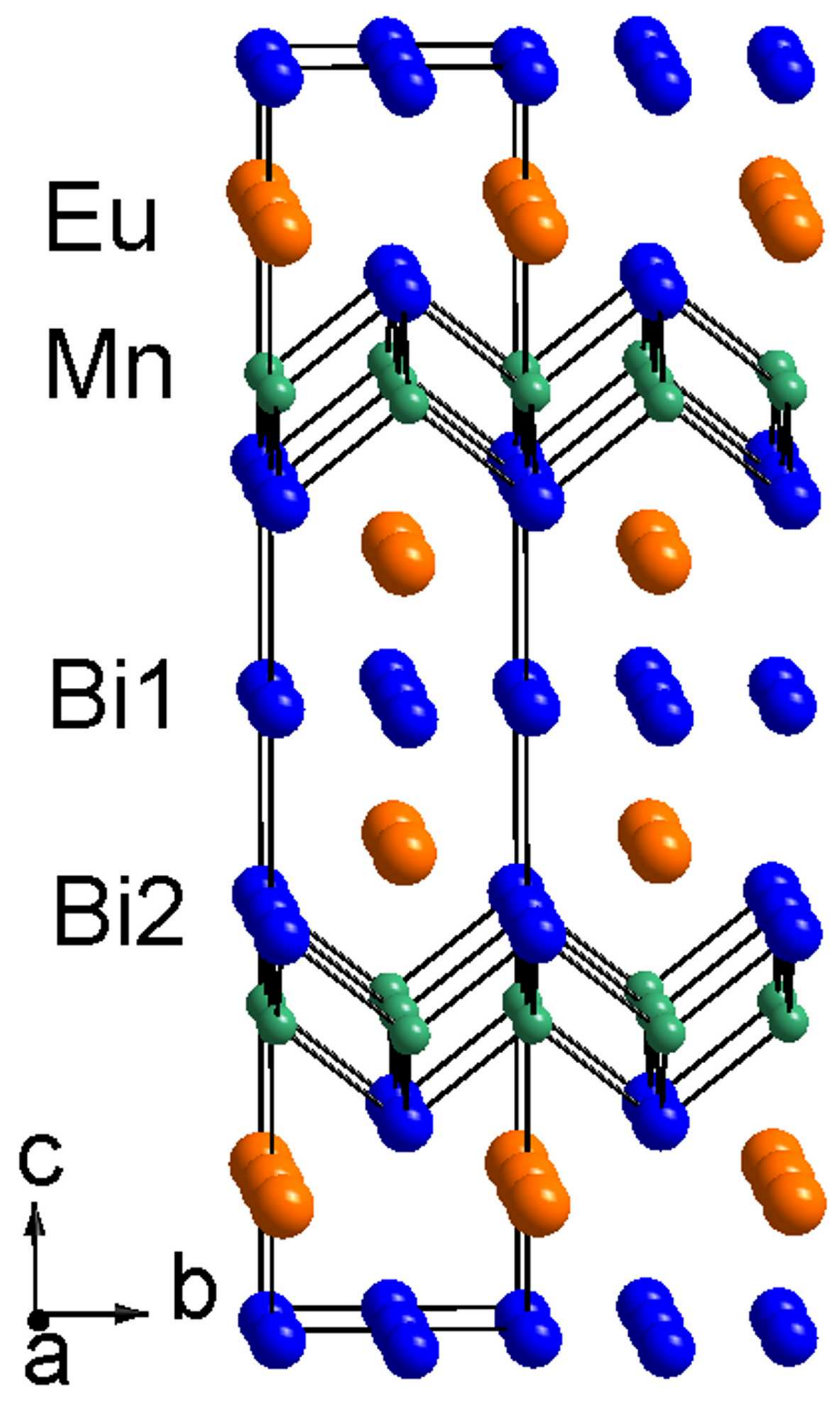}
    \end{minipage}
     &
     \begin{itemize}
      \item[] \textit{a} (\AA)
      \item[] \textit{c} (\AA)   
    \item[] Eu, position
    \item[] Mn, position
    \item[] Bi1, position
    \item[] Bi2, position
    \item[] Eu, U$_{\textrm{eq}}$
    \item[] Mn, U$_{\textrm{eq}}$
    \item[] Bi1, U$_{\textrm{eq}}$
    \item[] Bi2, U$_{\textrm{eq}}$
      \item[] R$_1$/wR$_2$ (all data)
     \end{itemize}
     &
     \begin{itemize}
      \item[] 4.5342(6)
      \item[] 22.427(4)   
    \item[] 0 0 0.11479(3)
    \item[] 0 $\frac{1}{2}$ $\frac{1}{4}$
    \item[] 0 $\frac{1}{2}$ 0
    \item[] 0 0 0.32872(2)
    \item[] 0.0066(2) 
    \item[] 0.0077(5)
    \item[] 0.0066(2)
    \item[] 0.0062(2)
      \item[] 0.0178/0.0420
     \end{itemize}
     \end{tabular}
     \end{table}

In-plane electrical and thermal transport, and specific heat measurements were performed in a Quantum Design Physical Property Measurement System (PPMS), while magnetization measurements were performed in a Quantum Design Magnetic Property Measurement System (MPMS), as well as with the AC Measurement Option in the PPMS to allow for application of larger magnetic fields.  Electrical contacts were made in standard four-point configurations using platinum wires connected to the samples through either spot-welding or Ag paste, while Ag epoxy was utilized during the thermal transport measurements to provide electrical, thermal, and mechanical contacts for use in the PPMS Thermal Transport Option.  The specific heat capacity was obtained using the standard Quantum Design analysis software, as well as with the dual-slope method provided within the heat capacity option.  The heat capacity data were collected using Apiezon N-grease below 200\,K and H-grease above 220\,K.  Heat capacity measurements were also performed with a magnetic field applied along the crystallographic $c$-axis (to within approximately 10$^{\circ}$).

The Hall coefficient, $R_H$, was obtained at fixed temperatures by sweeping the magnetic field (applied along the $c$-axis) and using the odd-in-\textbf{H} transverse resistance $\rho_{xy}=\frac{1}{2}[\rho(+\textbf{\textrm{H}})-\rho(-\textbf{\textrm{H}})$], which is particularly important in this material due to the large magnetoresistance.  Magnetoresistance measurements were performed using the AC Transport Option in the PPMS, with the rotator being employed to provide orientation-dependent information while a Cernox thermometer verified temperature stability. The measured magnetoresistance, defined as MR = $\frac{\rho(\textbf{H})-\rho(\textbf{H}=0)}{\rho(\textbf{H}=0)}$, often possessed an asymmetry between positive and negative values of the applied magnetic field \textbf{H}.  This effect was minimized by utilizing spot-welded voltage leads to minimize transverse displacements between leads, indicating the asymmetry is likely caused by a strong Hall contribution when a transverse offset between the voltage leads exists.  The reported magnetoresistance data deviate from the even-in-\textbf{H} contribution of the total measured resistance (expected absolute magnetoresistance) by 2\% at most, at 5\,K and 12\,T.

\section{Results and Discussion}

\subsection{Magnetization}

\begin{figure}[h!]%
\includegraphics[width=\columnwidth]{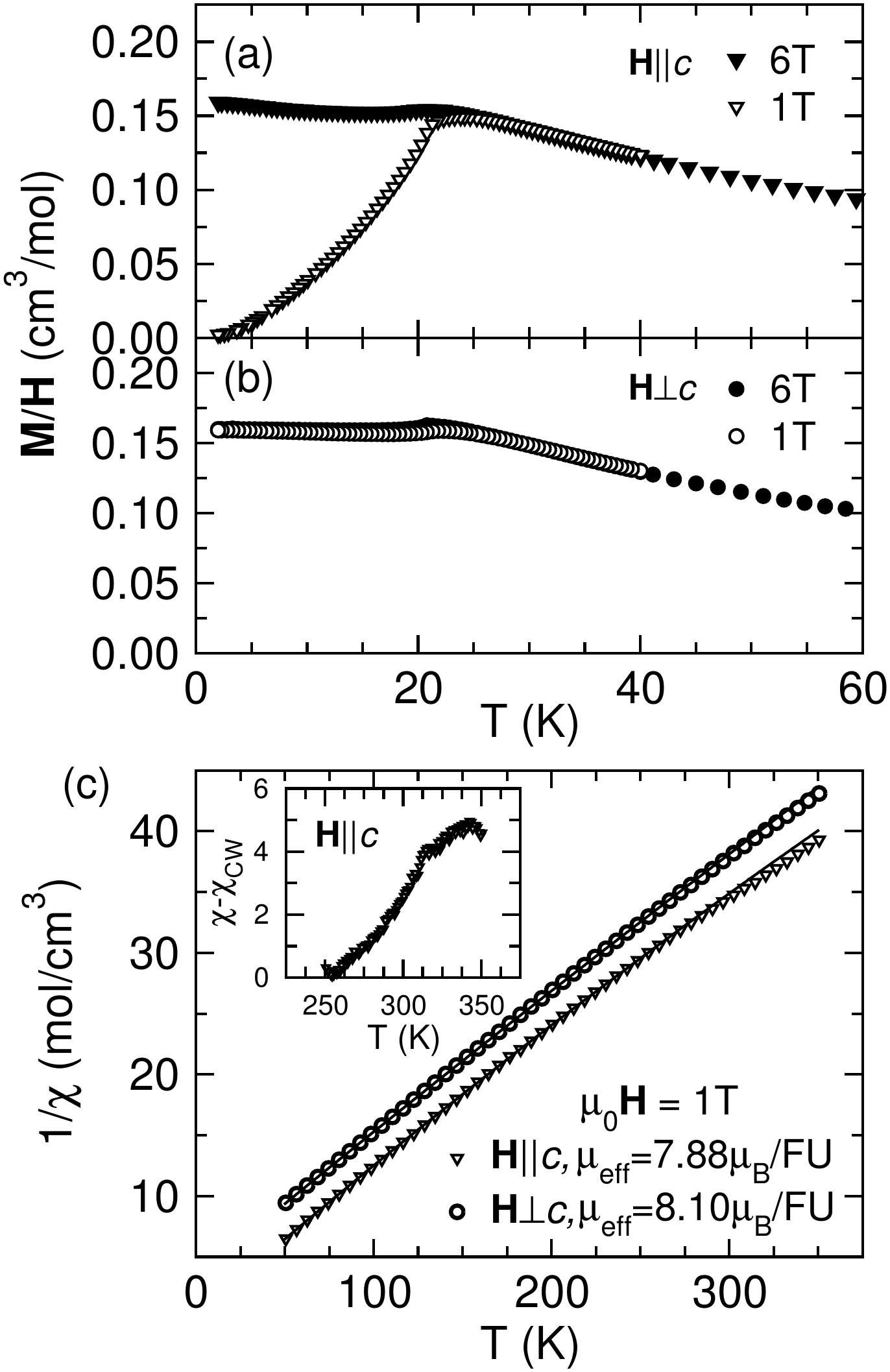}%
\caption{(a,b) Magnetic moment divided by applied field is plotted versus $T$ in the region of the antiferromagnetic ordering of Eu moments at $T_N$=22\,K; data collected in 1\,T reveals Eu moments align along $c$-axis while the 6\,T data set demonstrates the influence of a spin-flop in reducing the anisotropy in $\textbf{M/H}$. In panel (c), fits (solid lines) to modified Curie Weiss laws between 50 and 275\,K are shown for 1/$\chi=\textbf{H/M}$; data for \textbf{H}$\parallel c$ has been shifted by -2.5\,mol/cm$^3$ for clarity.  The inset in panel (c) shows the difference between the observed data and the Curie-Weiss model, thereby revealing the ordering of moments on the Mn sublattice near 315\,K.}%
\label{chi}%
\end{figure}

Temperature-dependent magnetization data for EuMnBi$_2$ are shown in Figure \ref{chi} for the two primary crystallographic directions (parallel and perpendicular to the $c$-axis).  Data for applied fields $\mu_0$\textbf{H} = 1\,T and 6\,T are shown, and these fields are above and below a spin-flop transition that is discussed in detail below.  The magnetization $\textbf{M}$ displays typical Curie-Weiss behavior from approximately 30\,K-300\,K.  This is demonstrated in Figure \ref{chi}c, where fits of the susceptibility ($\chi$=$\textbf{M/H}$) to a modified Curie-Weiss law are demonstrated above 50\,K (data from 50-275\,K are fit).  The data are fit to $\chi$= $\chi_0$ + $\frac{C}{T - \theta}$, where the Curie constant $C$ is related to the effective moment in the usual manner,\cite{Smart} and $\chi_0$ accounts for temperature independent contributions.  The effective moments obtained in this manner are generally consistent with those expected for divalent Eu ($\mu_{\textrm{eff}}$=7.94$\mu_{\textrm{B}}$) and the Weiss temperatures ($\theta$) are typically between -20 and -30\,K.  Analysis of an anomaly in specific heat is also consistent with divalent Eu, as discussed below.

In isostructural SrMnBi$_2$ and structurally similar CaMnBi$_2$, the Mn moments order antiferromagnetically near 270-290\,K with Mn moments parallel and anti-parallel to the $c$-axis.\cite{Park2011PRL,Wang2011,He2012APL}  It is therefore reasonable to expect similar behavior for Mn in EuMnBi$_2$.  However, the large paramagnetic contribution from Eu$^{2+}$ strongly masks the magnetic behavior of Mn, and an anomaly associated with Mn ordering can only be observed after analysis of the susceptibility data.  After the data for EuMnBi$_2$ are fit to a modified Curie-Weiss law, the residual allows the contribution of the Mn sublattice to be examined.  Through this analysis, a clear decrease in the residual $\chi$ can be observed below approximately 315\,K when $\textbf{H}\parallel c$-axis (see inset of Fig.\ref{chi}c).  This behavior is consistent with data reported for SrMnBi$_2$, where antiferromagnetic Mn ordering near 280\,K causes a decrease in $\chi$ below 280\,K when the field is applied along the $c$-axis.\cite{Park2011PRL}  We have confirmed behavior similar to that in Ref. \citenum{Park2011PRL} for $\chi(T)$ of our SrMnBi$_2$ crystals.  The anisotropy observed in the effective moments (Fig.\,\ref{chi}c) is also consistent with this antiferromagnetic order of Mn moments, which leads to a suppression of the calculated effective moment when data for $\textbf{H}$ $\parallel c$-axis are fit to a modified Curie-Weiss law.  By this comparison with SrMnBi$_2$, our magnetization data suggest Mn in EuMnBi$_2$ orders AFM at $\approx$315\,K with moments along the $c$-axis.  The slightly smaller lattice parameters (1\%-3\%) likely lead to the higher Mn ordering temperature in EuMnBi$_2$ compared to SrMnBi$_2$. 

\begin{figure}[h!]%
\includegraphics[width=0.8\columnwidth]{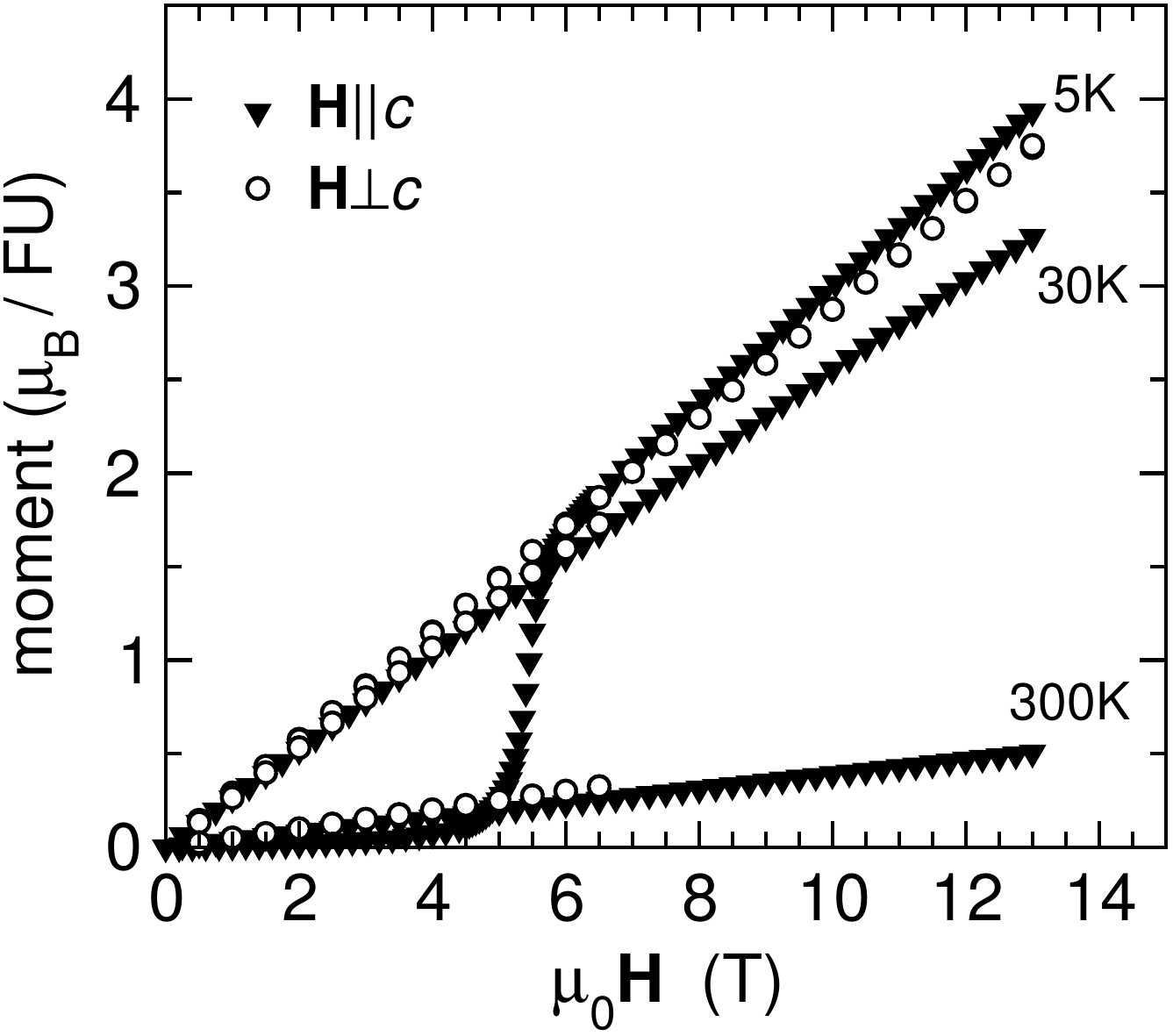}%
\caption{Field-dependence of the magnetic moment in EuMnBi$_2$ reveals a spin-flop transition for applied fields parallel to the \textit{c}-axis below $T_N$.}%
\label{MH}%
\end{figure}

The Eu moments order antiferromagnetically at $T_N$=22\,K, as demonstrated in Figure \ref{chi}(a,b).  For $\mu_0$\textbf{H}$\lesssim$5\,T, the temperature and orientation dependence of $\textbf{M/H}$ are consistent with Eu moments being aligned along the $c$-axis.  However, the magnetic response changes for $\mu_0$\textbf{H}$\gtrsim$5\,T, and this change in magnetization is best observed in plots of $\textbf{M}$ versus \textbf{H} at various $T$ (Fig.\ref{MH}).

A spin-flop transition is observed below $T_N$, with an onset near $\approx$5.2\,T and centered at $\approx$5.4\,T at 5\,K, as shown in Figure \ref{MH}.  The observed field-dependence is typical for a spin-flop transition in an antiferromagnet with spins aligned along $c$.\cite{Buschow}  This behavior suggests that the magnetocrystalline anisotropy is small relative to the AFM coupling energy but is large compared to other Eu$^{2+}$-containing compounds, which have quenched orbital angular momentum L=0.  If the magnetocrystalline anisotropy were very large, the spins would reorient directly into a saturated/fully-polarized state (a spin-flip).\cite{Buschow} Since the spin-flop is only observed below $T_N$, it seems likely that the Eu moments reorient. Interestingly, the spin-flop field of $\approx$5.2\,T corresponds to 24.5\,K for the divalent Eu ions with 7$\mu_B$, which is similar to $T_N$. The role of the Mn moments is unclear, though.  We are not aware of any similar spin-flop or spin-flip transitions being reported for Mn in CaMnBi$_2$ or SrMnBi$_2$, which is reasonable considering the high ordering $T$ of the Mn sublattice and the lower local moment.

The ordered moment does not saturate, even when fields of 13\,T are applied at 5\,K.  This is unusual for divalent Eu, which has a large local moment.  Typically, Eu moments order antiferromagnetically and become canted with application of a magnetic field, which leads an increase in the magnetization until a saturated moment of 7$\mu_B$/Eu is reached. The saturation field is usually rather modest (often on the order of a few tesla or less), and a slight anisotropy in this critical field is commonly observed in the antiferromagnetically ordered state for tetragonal or trigonal compounds. However, in cases where the moments are ordered antiferromagnetically along the $c$-axis, there is usually not a large anisotropy between the induced moments for \textbf{H}$\parallel$$c$ or \textbf{H}$\perp$$c$. Similarly, rapid changes in magnetization due to spin-flops or spin-flips are not commonly observed in divalent Eu compounds, particularly in tetragonal or trigonal examples containing pnictides. For instance, saturation is achieved at less than 5\,T in the trigonal EuZn$_2$Sb$_2$ and EuMg$_2$Bi$_2$ compounds,\cite{AMg2Bi2_Inorg,AZn2Sb2_2012} and an anisotropy on the order of 1\,T is observed in the saturation field but a spin-flop is not observed.  We note that a magnetic field of 13\,T corresponds to a temperature of 61\,K for Eu ions with 7$\mu_B$, and thus the lack of saturation at 13\,T and 5\,K is indeed surprising.

The high saturation field ($>$13\,T) at low $T$ is most likely caused by an interaction between the Eu and Mn moments.  Similarly, Eu-Mn interaction may enhance the magnetocrystalline anisotropy of Eu such that the spin-flop occurs.  Investigations of the response of the Mn moments to the ordering of the Eu moments, inelastic neutron scattering, and temperature-dependent crystallographic studies may provide additional insight into the coupling of these magnetic sublattices.

Relatively few materials with both Eu and Mn have been sufficiently characterized to provide a direct comparison.  EuMn$_2$Sb$_2$ is isostructural to EuZn$_2$Sb$_2$ (trigonal, CaAl$_2$Si$_2$ type) and the presence of Mn does lead to a complicated magnetic response.\cite{MossbauerEu122}  In this compound, Mn is reported to order ferromagnetically ($\sim$550-600\,K) and undergo a spin reorientation transition near 130\,K (based on specific heat data).  Below the antiferromagnetic ordering of Eu moments at 10\,K, the magnetization is almost linear up to approximately 4\,T, at which point saturation begins and a minor spin reorientation is observed near 6\,T before saturation is essentially achieved near 8\,T.\cite{MossbauerEu122}  While saturation was reached in EuMn$_2$Sb$_2$, the differences between it and EuMnBi$_2$ are rather large and the comparison only provides further support that interesting magnetic interactions exist in EuMnBi$_2$ and warrant further investigation.

\subsection{Specific Heat Capacity}

\begin{figure}[h!]%
\includegraphics[width=\columnwidth]{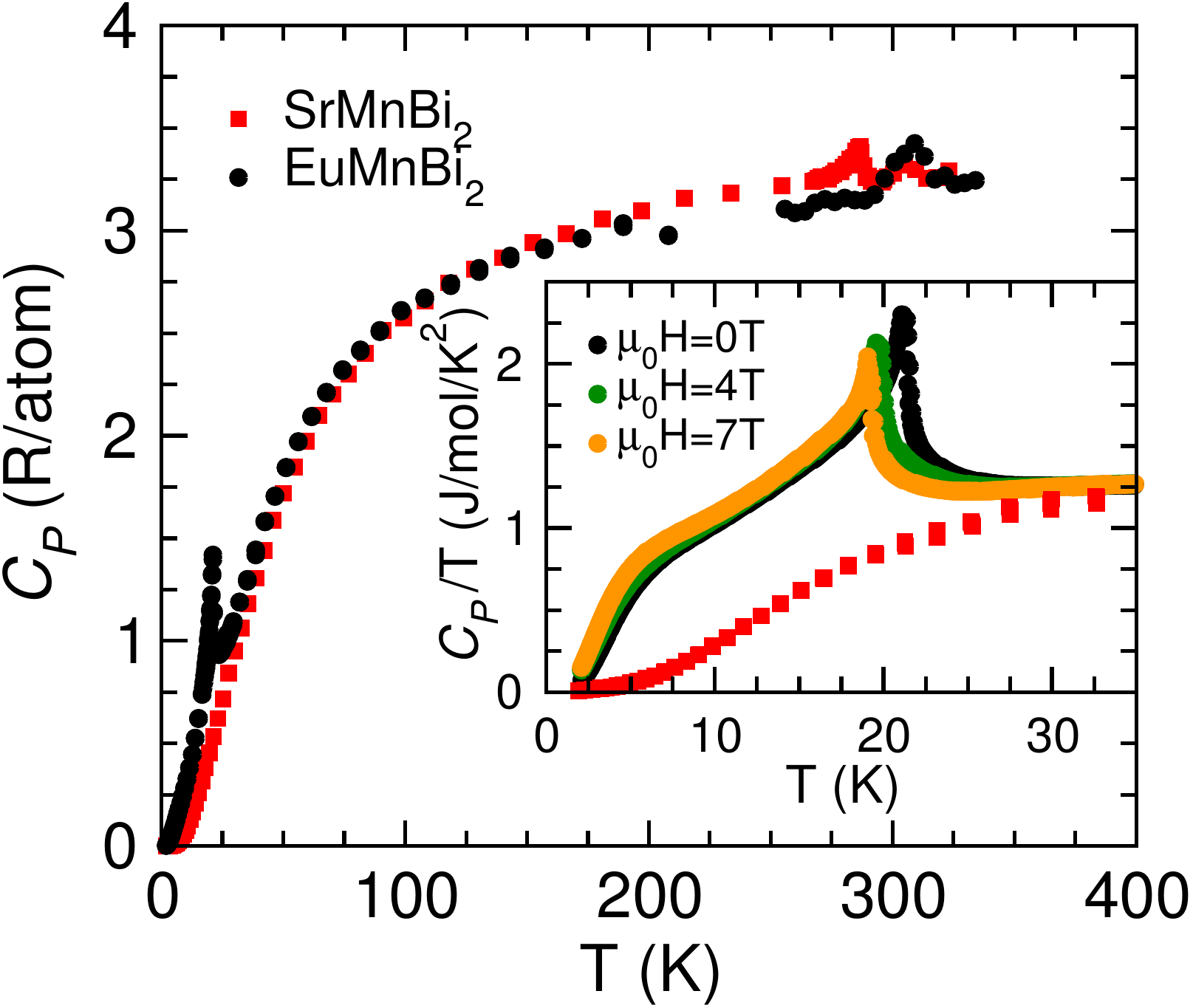}%
\caption{The specific heat capacity of EuMnBi$_2$ possesses anomalies near the ordering temperature of Eu moments at $T_N\approx$22\,K and near 307\,K, which is assumed to be due to ordering of Mn moments; data for SrMnBi$_2$ are also included.  Low-T data are emphasized in the inset as $C_P/T$, where the influence of a magnetic field is also shown.}%
\label{Cp}%
\end{figure}

The specific heat capacity $C_P$ of EuMnBi$_2$ possesses two anomalies in the temperature range of 2 to 330\,K, as shown in Figure\,\ref{Cp}.  A anomaly is observed near 307\,K, which is believed to be associated with the ordering of Mn moments.  In SrMnBi$_2$, the anomaly associated with Mn ordering has been reported to be rather weak,\cite{Wang2011} consistent with the results presented here for both EuMnBi$_2$ and SrMnBi$_2$.  This may suggest that there is in-plane (or short-range) ordering of Mn moments above the ordering temperatures obtained from magnetization measurements, which is reasonable due to the large distance along $c$ between Mn sites.

A strong anomaly in $C_P$ associated with the ordering of Eu moments is observed near 21\,K.  The low-$T$ anomaly is highlighted in the plot of $C_p/T$, which is shown in the inset of Fig.\,\ref{Cp}, and a shoulder is observed near 6\,K. The entropy change associated with the anomaly near $T_N$ for EuMnBi$_2$ is determined by integrating $C_P/T$, using scaled data for SrMnBi$_2$ as a baseline.  The integration yields values consistent for ordering of paramagnetic moments on divalent Eu with J=$\frac{7}{2}$, where $\Delta$S=Rln(2J+1)=17.3\,J/mol/K. Specifically, integrating from 2-35\,K yields 17.2\,J\,/mol/K, which also includes any contribution of the low-$T$ shoulder.  These values are reported using the data shown in the inset of Fig.\,\ref{Cp}, which were obtained using the ``dual-slope method'' provided by Quantum Design.  A similar value of 16.1\,J/mol/K was obtained using data from another measurement (different crystal) that employed the more common small heat pulse (3\% temperature rise) with the default analysis from the Quantum Design Heat Capacity Option. 

Application of a magnetic field has a slight influence on the low-$T$ heat capacity anomaly.  The peak position shifts to lower temperature with increasing magnetic field, consistent with an antiferromagnetic ordering transition.  The shoulder near 6\,K displays less magnetic field dependence, though.  This may suggest that the low-$T$ shoulder is caused by a structural feature.  However, the calculated $\Delta$S values are all similar and approximately equal to those expected for the magnetic transition alone. For instance, in applied fields of 4\,T and 7\,T we find $\Delta$S=17.1\,J/mol/K and 16.8\,J/mol/K, respectively.  Thus, it is not possible to isolate the role or origin of the low-$T$ shoulder based upon the current results.

\subsection{Electrical and Thermal Transport}

\begin{figure}[hb!]%
\includegraphics[width=\columnwidth]{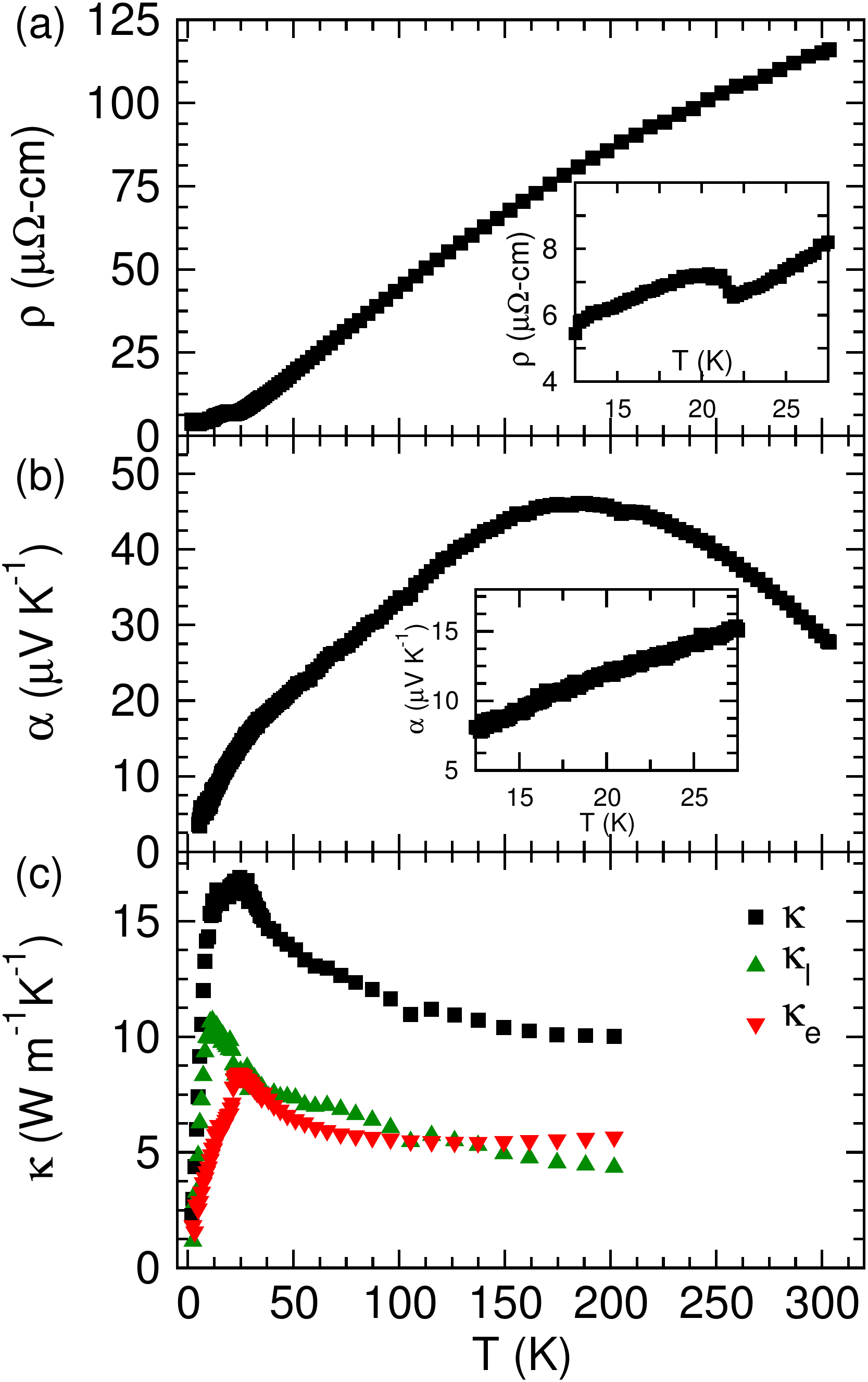}%
\caption{In-plane electrical resistivity (a), Seebeck coefficient (b) and thermal conductivity (c) of EuMnBi$_2$, with insets emphasizing the behavior near $T_N\approx$22\,K.}%
\label{elec}%
\end{figure}

EuMnBi$_2$ displays transport properties typical of a semimetal or lightly-doped, narrow gap semiconductor.  As shown in Figure \ref{elec}(a), the electrical resistivity $\rho$ increases with increasing temperature, and the increase is nearly linear with $T$ up to about 100-150\,K before slowing down. A small increase in $\rho$ is observed on cooling below $T_N$, which is explained by the Hall data discussed below.

From the lowest temperatures, the Seebeck coefficient $\alpha$ is positive and increases with increasing $T$, suggesting holes dominate conduction. A maximum in $\alpha$ is observed near $\approx$180\,K, above which temperature the magnitude of $\alpha$ decreases with increasing $T$ (Fig.\ref{elec}(b)).  This suggests the addition of free carriers with increasing $T$, which may be caused by the activation of electron-hole pairs across a small gap, the activation of additional holes from impurity states, or the broadening of the Fermi distribution function if a negative or zero band gap exists in the electronic structure.  Similar data have been reported for SrMnBi$_2$ and CaMnBi$_2$, where a strong magnetic field influence was observed.\cite{Wang2012APL}  We do not observe the pronounced low-$T$ hump in $\alpha$ near 50\,K that was attributed to phonon drag.\cite{Wang2012APL}

The thermal conductivity $\kappa$ of EuMnBi$_2$ is shown in Figure \ref{elec}(c).  $\kappa$ generally behaves as expected for a crystalline material, with a decay in the lattice contribution $\kappa_l$ above $\approx$10\,K being observed.  As expected from the $\rho$ data, the electronic contribution $\kappa_{e}$ decreases below $T_N$, and this causes the complex shape of the measured $\kappa$.  The lattice and electronic contribution were obtained assuming the degenerate limit of the Lorenz number within the typical Wiedemann-Franz expression for $\kappa_{ele}$. This assumption is reasonable given the metallic nature of EuMnBi$_2$, though the thermal activation of carriers limits its validity at the higher temperatures examined.

\subsection{Hall effect}

\begin{figure}[hb!]%
\includegraphics[width=\columnwidth]{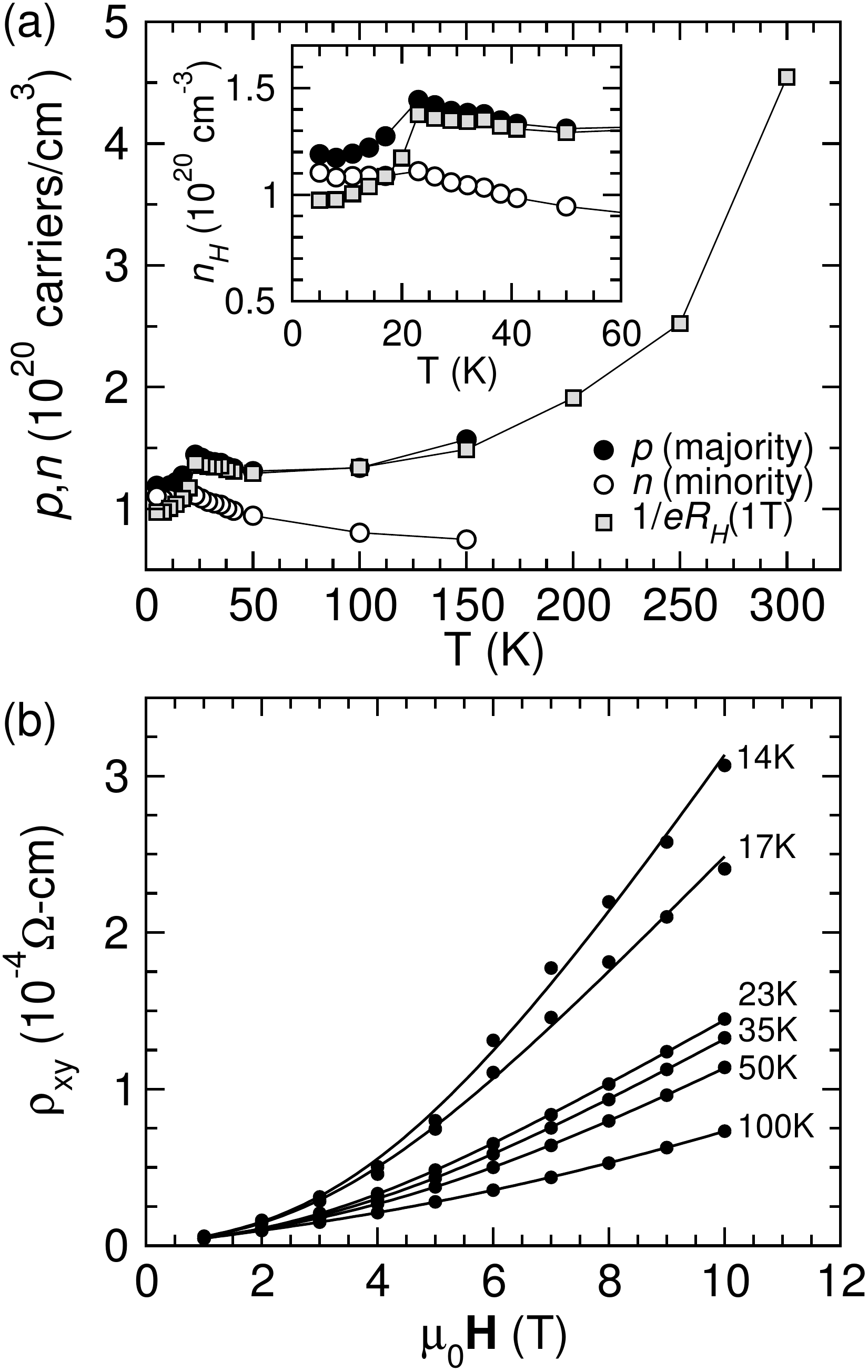}%
\caption{(a) Hole $p$ and electron $n$ carrier densities obtained by fitting the Hall resistance $\rho_{xy}$ in panel (b) to a two carrier model (solid lines) at various temperatures.  Also shown in (a) is the hole carrier density calculated using a single carrier model with the Hall coefficient $R_H$ obtained at an applied field of 1\,T, which displays similar quantitative and qualitative behavior to the hole density obtained from the two carrier model.}%
\label{Hall}%
\end{figure}

The Hall effect can provide a fine probe of the electronic structure in a material.  The Hall coefficient is given by $R_H=\rho_{xy}/\textbf{\textrm{H}}$, where $\rho_{xy}$ is the Hall resistance obtained from the odd-in-\textbf{\textrm{H}} part of the measured (transverse) resistance.  In EuMnBi$_2$, the Hall coefficient reveals a decrease in the hole carrier density $p$ below the magnetic ordering of Eu at $T_N$=22\,K, as observed in Figure \ref{Hall}a.  This demonstrates a direct interaction between magnetism on Eu sites and the Fermi surface.  Furthermore, the Hall coefficient is independent of magnetic field for $T\gtrsim\,150\,K$, but becomes strongly field-dependent at low temperatures.  This suggests multiple bands/carrier types influence $R_H$, particularly at low $T$, which is observed as the non-linear dependence of $\rho_{xy}$ on applied field shown in Figure \ref{Hall}b.

The field-dependence of $R_{H}$ can be understood in terms of a multi-carrier model, with dominant conduction by holes and a minority contribution from low-mobility electrons.  Based on  calculations reported for the electronic structure of isostructural SrMnBi$_2$,\cite{Lee2013PRB} we expect at least one hole and one electron band to contribute to conduction.  To maintain the simplest model possible, we have analyzed the field-dependent Hall data with a two-band model.

For a two-band system, the field-dependence of the Hall coefficient is determined by the conductivity $\sigma_i$ and Hall coefficient $R_i$ of each band $i$:\cite{Putley}

\begin{equation}
R_H=\frac{R_1\sigma_1^2+R_2\sigma_2^2+H^2R_1R_2\sigma_1^2\sigma_2^2(R_1+R_2)}{(\sigma_1+\sigma_2)^2+H^2\sigma_1^2\sigma_2^2(R_1+R_2)^2}
\label{Hall2Band}
\end{equation}

\noindent Here, we use $i=p$ for holes and $i=n$ for electrons ($R_n<0$).  As shown in Figure \ref{Hall}b, the data are well described by Eq. \ref{Hall2Band} below 150\,K, with the quality of fit degrading on cooling below $T_N$ due to the spin-flop transition near 5\,T. As shown in Figure \ref{Hall}a, the fit parameters reveal that the hole concentration $p=1/eR_p$ is very similar to that obtained using the absolute Hall coefficient at 1\,T (single carrier assumption).  This is due to the relatively small contribution of electrons to the total conductivity; the fitted conductivities are observed to obey $\sigma_p/\sigma>$99.8\% for all temperatures examined.  The mobility of holes increases from $\approx$90\,cm$^2$/V/s at 150\,K to $\approx$520\,cm$^2$/V/s at 5\,K, while that of electrons changes from only 0.12\,cm$^2$/V/s to 1.6\,cm$^2$/V/s between the same temperatures.

The fittings were performed using the total conductivity ($\sigma_h=\sigma(H)-\sigma_n$) as a constraint to eliminate one free parameter.  To obtain reasonable initial guesses and upper/lower limits for the remaining three free parameters ($R_p$,$R_n$,$\sigma_n$), the data at 23\,K were initially fit using the low-field limit of Eq. \ref{Hall2Band} as an additional constraint.  These results were not directly used to initialize the three free parameters, though the final solution(s) was similar to these values.  No other local minima were observed during the fitting routines.  However, the limits of Eq. \ref{Hall2Band} should be recognized.  In particular, it does not account for field dependent $\sigma_i$ or $R_i$.  The spin-flop may also produce a change in the individual Hall coefficients, leading to degradation of the fit at low temperatures and high fields.  We note that similar results (within 5\%) were obtained at 50\,K when the field-dependent conductivity was used for the initial constraint $\sigma_h=\sigma(\textbf{\textrm{H}})-\sigma_n$.  Similar $R_H$ results were observed for various crystals from different growths.

The Hall effect data are generally consistent with the behavior of $\rho$ shown above. In particular, the decrease in hole concentration corresponds to the increase in resistivity below $T_N$. This is reasonable, as the Eu-Bi layer is expected to influence the bands that dominate conduction (those from the Dirac cones, in analogy to SrMnBi$_2$).\cite{Wang2011,Lee2013PRB}  Interestingly, this transition does not influence the temperature dependence of $\alpha$ (inset, Fig\ref{elec}b), which tends to be sensitive to such electronic changes.  In any case, there is a clear influence of Eu magnetism on the electronic properties, and placing magnetic moments on the Bi nets' nearest-neighbors may provide an avenue to tune the nature of the Dirac cones.

\subsection{Magnetoresistance}

\begin{figure}[h!]%
\includegraphics[width=\columnwidth]{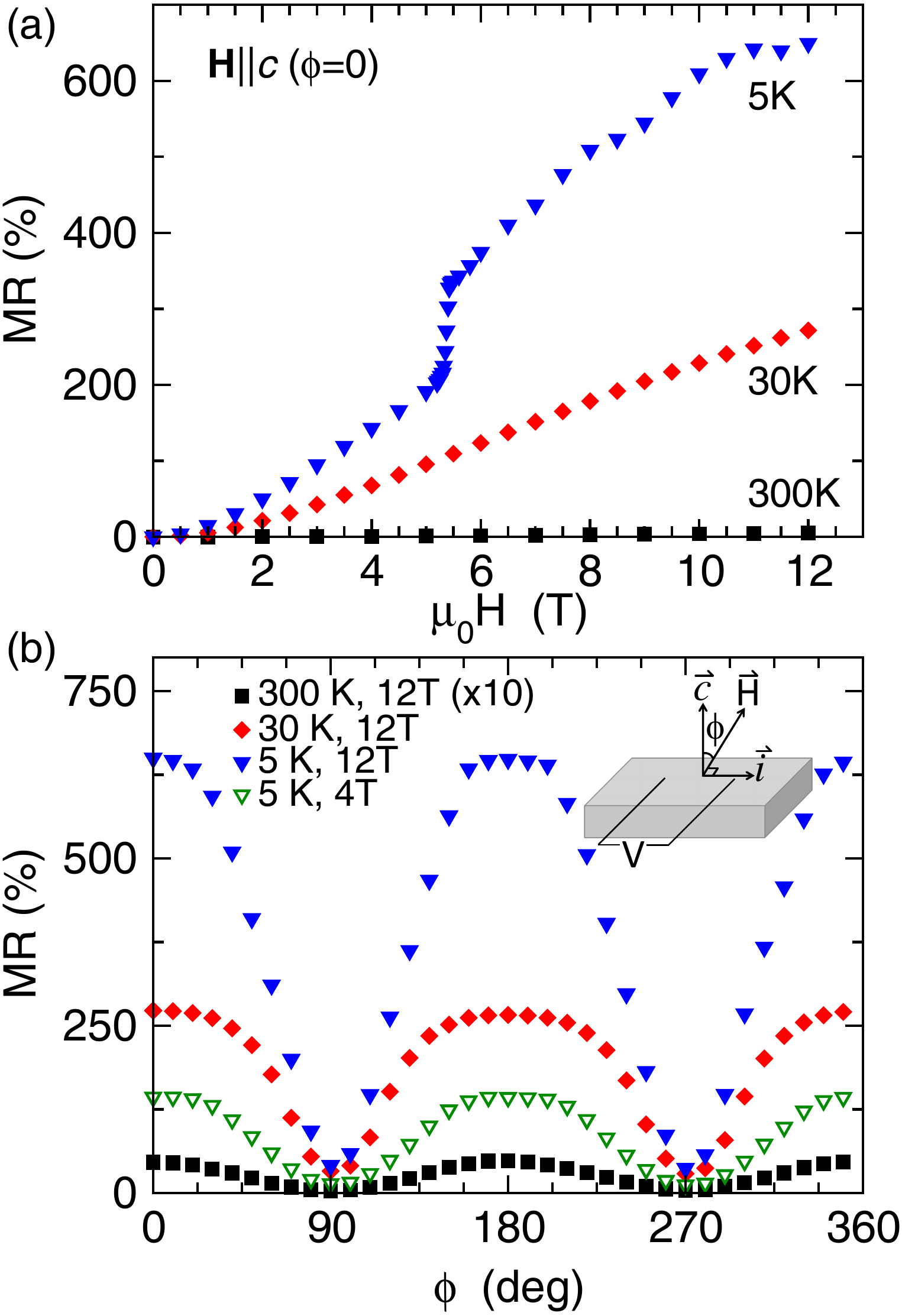}%
\caption{Magnetoresistance as a function of (a) applied field and (b) angle of applied field with respect to the $c$-axis.  In (b), the data collected at 300\,K have been multiplied by a factor of 10.}%
\label{MR}%
\end{figure}

The magnetoresistance (Fig. \ref{MR}) is perhaps the most striking physical property of EuMnBi$_2$.  Like isostructural SrMnBi$_2$ and electronically/structurally-similar CaMnBi$_2$ and LaAgBi$_2$, the magnetoresistance is intrinsically large and positive in EuMnBi$_2$.  The magnetoresistance increases linearly with increasing field, which is expected to occur above a critical field in these compounds due to linear band dispersions associated with Dirac points near the Fermi surface.\cite{Wang2013PRB,Abrikosov2003}

The magnetoresistance (MR) in EuMnBi$_2$ is somewhat unique, though, because a large enhancement is observed below $T_N$ and across the spin-flop transition.  For the data shown, the MR increases from $\approx$200\% to $\approx$300\% between 5.2 and 5.4\,T (at 5\,K).  It is difficult to determine if this sharp increase in MR is due to a reconstruction of the Fermi surfaces or simply a decrease in carrier scattering due to the reorientation of the moments.  The Hall data clearly suggest a decrease in the carrier density due to the ordering of Eu moments, and thus it is reasonable to suspect that the spin-flop may also influence the Fermi surface.


At the highest applied fields, non-linear magnetoresistance is observed. This behavior, qualitatively present in all samples, can be ascribed to quantum oscillations.  Shubnikov-de Haas and De Hass-van Alphen oscillations are clearly observed in SrMnBi$_2$, CaMnBi$_2$, LaAgSb$_2$ and LaAgBi$_2$, where analysis of these oscillations is used to verify the presence of Dirac fermions.\cite{Park2011PRL,Wang2012PRB,Wang2012APL}

The magnetoresistance is highly anisotropic, revealing the two-dimensional nature of EuMnBi$_2$.  This anisotropy is highlighted in Figure \ref{MR}(b), where the magnetoresistance is plotted as a function of the direction ($\phi$) of the applied field relative to the $c$-axis ($\vec{c}$).  $\phi$=0 is the standard longitudinal magnetoresistance, where magnetoresistance is maximized in EuMnBi$_2$.  The magnetoresistance is minimized when the field is applied in the $ab$-plane.  This reveals the two-dimensional nature of the electronic transport in EuMnBi$_2$, and similar behavior is observed for fields above and below the spin-flop transition.  Classically, in the $\phi$=90$^{\circ}$ orientation the magnetoresistance is influenced by the mobility of carriers along the $c$-axis (a component of the plane perpendicular to the applied field).  As such, this measurement reveals that the carrier transport is anisotropic (highly 2D) with a low effective carrier mobility along the $c$-axis.  Similar anisotropy has been reported for CaMnBi$_2$ and LaAgBi$_2$.\cite{Wang2012APL,Lee2013PRB}  Interestingly, at 5\,K and $\phi$=90$^{\circ}$, the field dependence of the MR changes at the critical field associated with the spin-flop; we observed MR $\propto\textbf{\textrm{H}}^2$ below 5\,T and MR linear with field above 5.5\,T for $\phi$=90$^{\circ}$.

\section{Summary}

EuMnBi$_2$ is a layered compound that is structurally similar to other ternary pnictides that have become of interest because they possess linear bands dispersing from Dirac cones near the Fermi surface.  These linear bands produce large, positive magnetoresistance that increases linearly with increasing field and does not saturate. One feature that separates EuMnBi$_2$ from these other compounds is antiferromagnetic ordering of large moments on divalent Eu ions at $T_N$=22\,K.  The magnetic ordering of Eu moments leads a suppression of the carrier density, and thus demonstrates the influence of magnetism on the Fermi surface.  Below $T_N$, a spin-flop occurs for magnetic fields of approximately $\approx$5.4\,T applied along the $c$-axis, and a correspondingly sharp increase in the magnetoresistance is also observed.  Interestingly, the induced magnetization is not saturated in an applied field of 13\,T at 5\,K.  This is most likely related to an interaction between localized moments on Mn and Eu, the former of which appear to order antiferromagnetically near approximately 310\,K.  Similarly, such an interaction likely promotes magnetocrystalline anisotropy leading to the spin-flop transition and the associated increase in magnetoresistance.  As such, isolating the interactions between the ordered moments should be one primary goal of further studies.

\section{Acknowledgements}

This work was supported by the U. S. Department of Energy, Office of Basic Energy Sciences, Materials Sciences and Engineering Division.  We thank Radu Custelcean for assistance with single crystal x-ray diffraction measurements.

\end{document}